\documentclass[aps,prd,amssymb,nofootinbib,twocolumn,epsf,floatfix,showpacs]
{revtex4}
\usepackage[usenames]{color} 
\usepackage{ulem} 
\usepackage{graphicx}
\usepackage{amssymb}
\usepackage{amsmath}
\usepackage{epstopdf}

\begin{document}


\title{Improved Spectral Representations of Neutron-Star
  Equations of State}

\author{Lee Lindblom}

\affiliation{Center for Astrophysics and Space Sciences, University 
of California at San Diego, La Jolla, CA 92093, USA}

\date{\today}
 
\begin{abstract}
  Spectral representations have been shown to provide an efficient way
  to represent the poorly understood high-density portion of the
  neutron-star equation of state.  This paper shows how the efficiency
  and accuracy of those representations can be improved by a very
  simple change.
\end{abstract}


\maketitle


\section{Introduction}
\label{s:introduction}

Spectral representations have been shown to be accurate and efficient
ways to parameterize the high-density portions of the neutron-star
equation of state~\cite{Lindblom2010, Lindblom2018}.  Consequently
these representations are being used by several groups now to extract
equation of state information from data obtained by x-ray and
gravitational-wave observations of neutron stars, see
e.g.~\cite{abbott2018gw170817, miller2019psr,
  raaijmakers2020constraining}.  This paper shows how a very simple
change can increase the accuracy and efficiency of these spectral
representations.

Section~\ref{e:BasicRepresentations} briefly reviews and then shows
how to improve the basic pressure-based representations constructed
from spectral expansions of the adiabatic index of the material,
see~\cite{Lindblom2010}.  Section~\ref{e:CausalRepresentations}
reviews and then shows how to improve the pressure-based causal
representations constructed from spectral expansions related to the
sound-speed of the material, see~\cite{Lindblom2018}.  Finally,
Sec.~\ref{s:EnthalpyRepresentations} shows how these improvements can
be extended to the enthalpy-based representations of the equations of
state, which are useful for some purposes~\cite{Lindblom1992}.  The
accuracy and efficiency of the improved spectral representations in
each section are evaluated by computing and then testing best-fit
representations of 27 realistic theoretical neutron-star equations of
state.

\section{Basic Spectral Representations}
\label{e:BasicRepresentations}

Previous studies~\cite{Lindblom2010, Lindblom2018} have shown that
efficient and accurate representations of neutron-star equations of
state can be constructed using spectral representations of key
thermodynamic quantities.  The first of these were based on spectral
expansions of the adiabatic index $\Gamma(p)$, defined by
\begin{equation}
\Gamma(p)=\frac{\epsilon(p)+p}{p}\left(\frac{d\epsilon}{dp}\right)^{-1}.  
\end{equation}
These basic spectral expansions have the form
\begin{equation}
  \Gamma(p)=\exp\left[\sum_k\gamma_k\Phi_k(p)\right],
  \label{e:Gamma}
\end{equation}
where the $\Phi_k(p)$ are a suitable fixed set of spectral basis
functions, and the $\gamma_k$ are the spectral coefficients that
determine a particular $\Gamma(p)$ in this representation.

As shown in~\cite{Lindblom2010}, any adiabatic index $\Gamma(p)$ on
the domain $p\geq p_0$ determines the equation of state $\epsilon(p)$
on that domain:
\begin{equation}
  \epsilon(p)=\frac{\epsilon_0}{\mu(p)}+\frac{1}{\mu(p)}\int_{p_0}^p
  \frac{\mu(p')}{\Gamma(p')}dp',
  \label{e:epsilon_p}
\end{equation}
where $\mu(p)$ is defined by
\begin{equation}
  \mu(p)=\exp\left[-\int_{p_0}^p \frac{dp'}{p'\Gamma(p')}\right].
  \label{e:mu_p}
\end{equation}
The spectral coefficients $\gamma_k$ determine the adiabatic index
from Eq.~(\ref{e:Gamma}), and then Eqs.~(\ref{e:epsilon_p}) and
(\ref{e:mu_p}) together with the integration constants $\epsilon_0$
and $p_0$ uniquely determine the equation of state in these basic
pressure-based spectral representations.

The constants $\epsilon_0$ and $p_0$ are the pressure and density at
the point where the spectral representation matches onto a low-density
equation of state: $\epsilon(p_0)=\epsilon_0$.  These constants,
$\epsilon_0$ and $p_0$, ensure that the match between the low-density
equation of state and the high-density spectral expansion is
continuous.  However, they do not ensure that the derivative of the
equation of state is continuous at this point.  The presence of a
discontinuity in the derivative would indicate, implicitly, the
presence of a phase transition in the material.  The matching point is
typically chosen at sufficiently low density that no physical phase
transition should be present there.  Additional constraints on the
spectral representation would be needed to ensure that no such
unphysical discontinuity occurs there.

The spectral basis functions used in the basic spectral expansions
introduced in~\cite{Lindblom2010} have the form,
\begin{equation}
  \Phi_k(p)=\left[\log\left(\frac{p}{p_0}\right)\right]^k.
  \label{e:p-basis-def}
\end{equation}
In the basic spectral expansions $\Gamma(p_0)=\exp(\gamma_0)$ since
$\Phi_0(p_0)=1$ and $\Phi_k(p_0)=0$ for $k>0$.  It follows that the
continuity at the matching point of $\Gamma(p)$, and hence the
derivative of the equation of state, is determined completely by the
lowest-order spectral coefficient $\gamma_0$.  This continuity is
assured if and only if $\gamma_0$ is chosen to be
$\gamma_0=\log\Gamma_0$, where $\Gamma_0$ is determined from the
low-density equation of state:
\begin{equation}
  \Gamma_0=\frac{\epsilon_0+p_0}{p_0}
  \left(\left.\frac{d\epsilon(p)}{dp}\right|_{p\uparrow p_0}\right)^{-1}.
  \label{e:Gamma0Def}
\end{equation}
\begin{figure}[t]
\centerline{\includegraphics[width=3in]{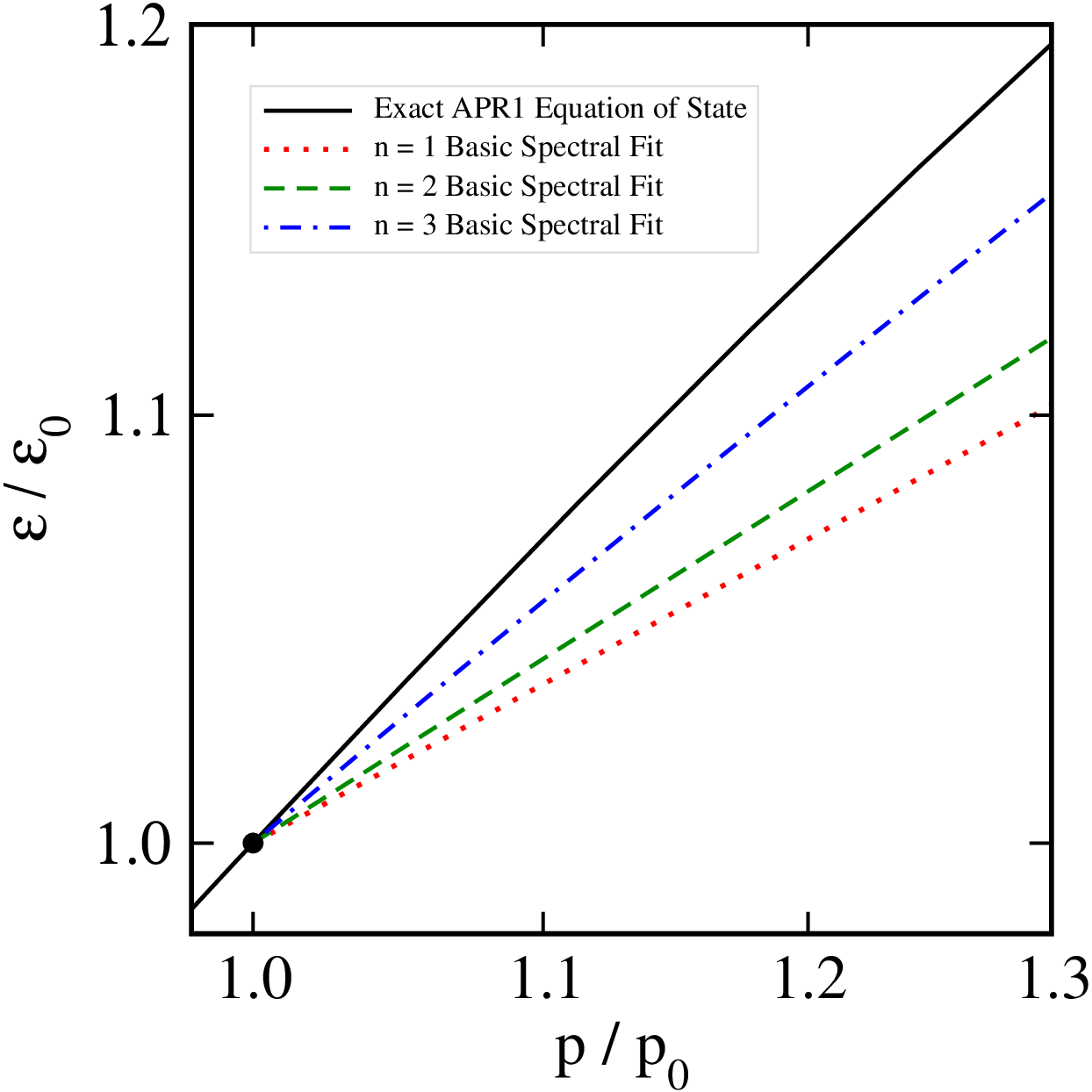}}
\caption{\label{f:spectral_apr1} Discontinuities in the slopes of the
  best-fit basic pressure-based spectral representations of the
  equation of state are illustrated at the point $(p_0,\epsilon_0)$
  (black dot) where they match onto the low density equation of state.
  The exact equation of state shown here as the solid (black) curve is
  the APR1 equation of state.  The (red) dotted curve is the best-fit
  $n=1$ parameter basic spectral equation of state, the (green) dashed
  curve is the best-fit $n=2$ curve, and the (blue) dot-dash curve is
  the best-fit $n=3$ curve.}
\end{figure}

In~\cite{Lindblom2010} the accuracy and efficiency of these spectral
representations were tested by adjusting the $n$ lowest spectral
coefficients $\{\gamma_0,\gamma_1, ..., \gamma_{n-1}\}$ to provided
the best-fit representations of realistic equation of state models in
the density range relevant for neutron stars. For finite $n$ the
spectral representations do not necessarily satisfy the derivative
continuity condition, $\Gamma(p_0)=\exp(\gamma_0)=\Gamma_0$, exactly.
(Although in the limit of large $n$ this discontinuity becomes
arbitrarily small in these representations.)
Figure~\ref{f:spectral_apr1} illustrates the slope discontinuities at
the matching point for the lowest-order basic spectral fits of the
APR1 equation of state.\footnote{The APR1 equation of state was chosen
as a representative because its spectral fits have errors that are
close to the averages of the errors for the model equations of state
studied in~\cite{Lindblom2010}.}

\begin{figure}[t]
\centerline{\includegraphics[width=3in]{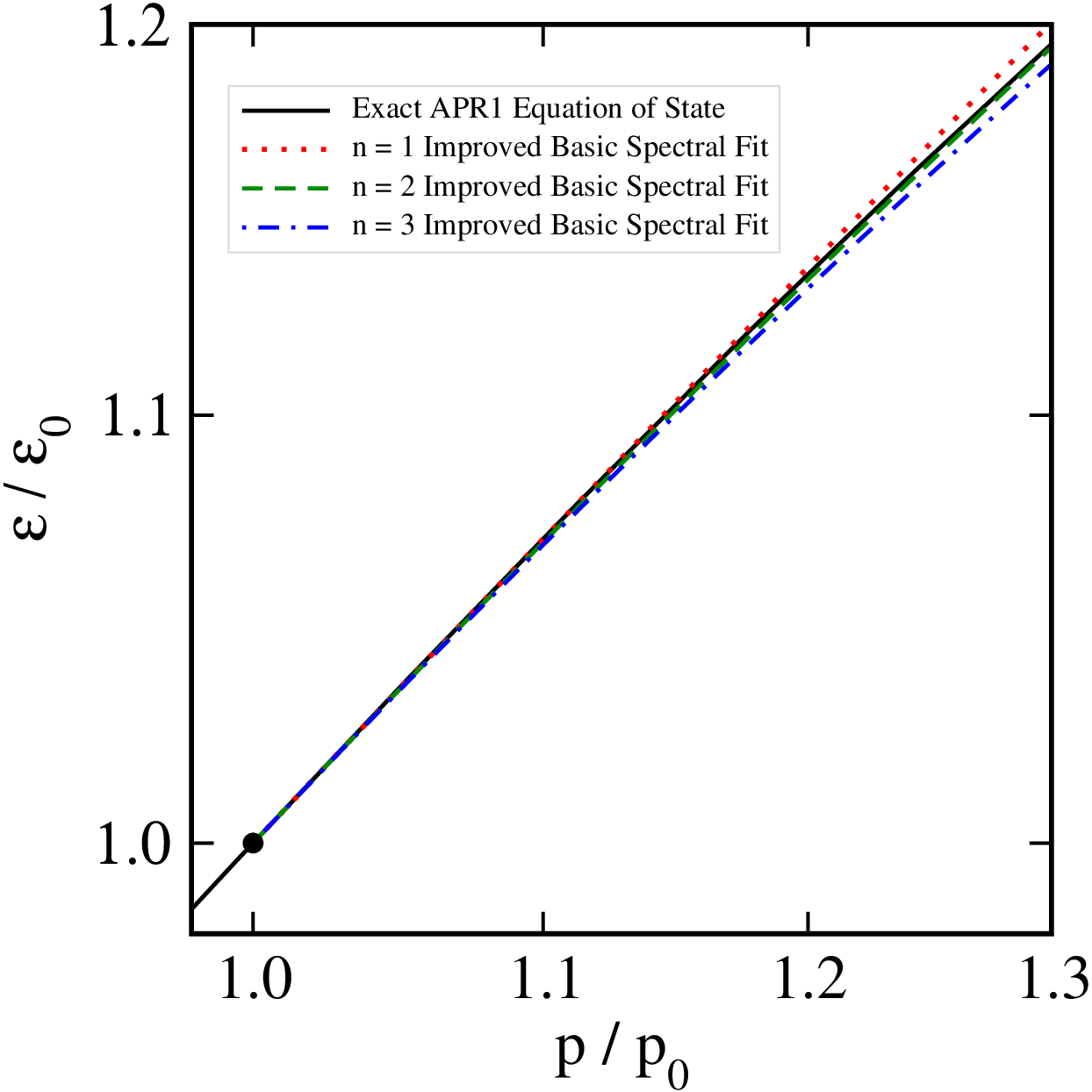}}
\caption{\label{f:new_p_spectral_apr1} Continuity and
  differentiability of the improved pressure-based basic spectral
  representations are illustrated at the point $(p_0,\epsilon_0)$
  (black dot) where the best-fit spectral equations of state match
  onto the low-density equation of state.  The exact equation of state
  shown here as the solid (black) curve is the APR1 equation of state.
  The (red) dotted curve is the best-fit $n=1$ parameter improved
  basic spectral equation of state, the (green) dashed curve is the
  best-fit $n=2$ curve, and the (blue) dot-dash curve is the best-fit
  $n=3$ curve.}
\end{figure}

The unphysical discontinuities in the derivatives of the equations of
state at the point $(p_0,\epsilon_0)$ can, however, be removed simply
by changing the way the best-fit spectral coefficients are
determined.  Instead of adjusting the spectral parameters
$\{\gamma_0,\gamma_1,...,\gamma_{n-1}\}$ in an $n$-parameter fit,
simply set
\begin{equation}
  \gamma_0=\log\Gamma_0,
  \label{e:gamm0def}
\end{equation}
where $\Gamma_0$ is defined in Eq.~(\ref{e:Gamma0Def}), and then
adjust the $n$ spectral coefficients
$\{\gamma_1,\gamma_2,...,\gamma_{n}\}$ instead.
Figure~\ref{f:new_p_spectral_apr1} illustrates the resulting best-fit
representations of the APR1 equation of state in the neighborhood of
the matching point $(p_0,\epsilon_0)$.  Comparing the improved
representations in Fig.~\ref{f:new_p_spectral_apr1} with the basic
representations in Fig.~\ref{f:spectral_apr1} shows that this simple
change is effective in removing the unphysical slope discontinuities
at the match point.
\begin{figure}[!htb]
\centerline{\includegraphics[width=3in]{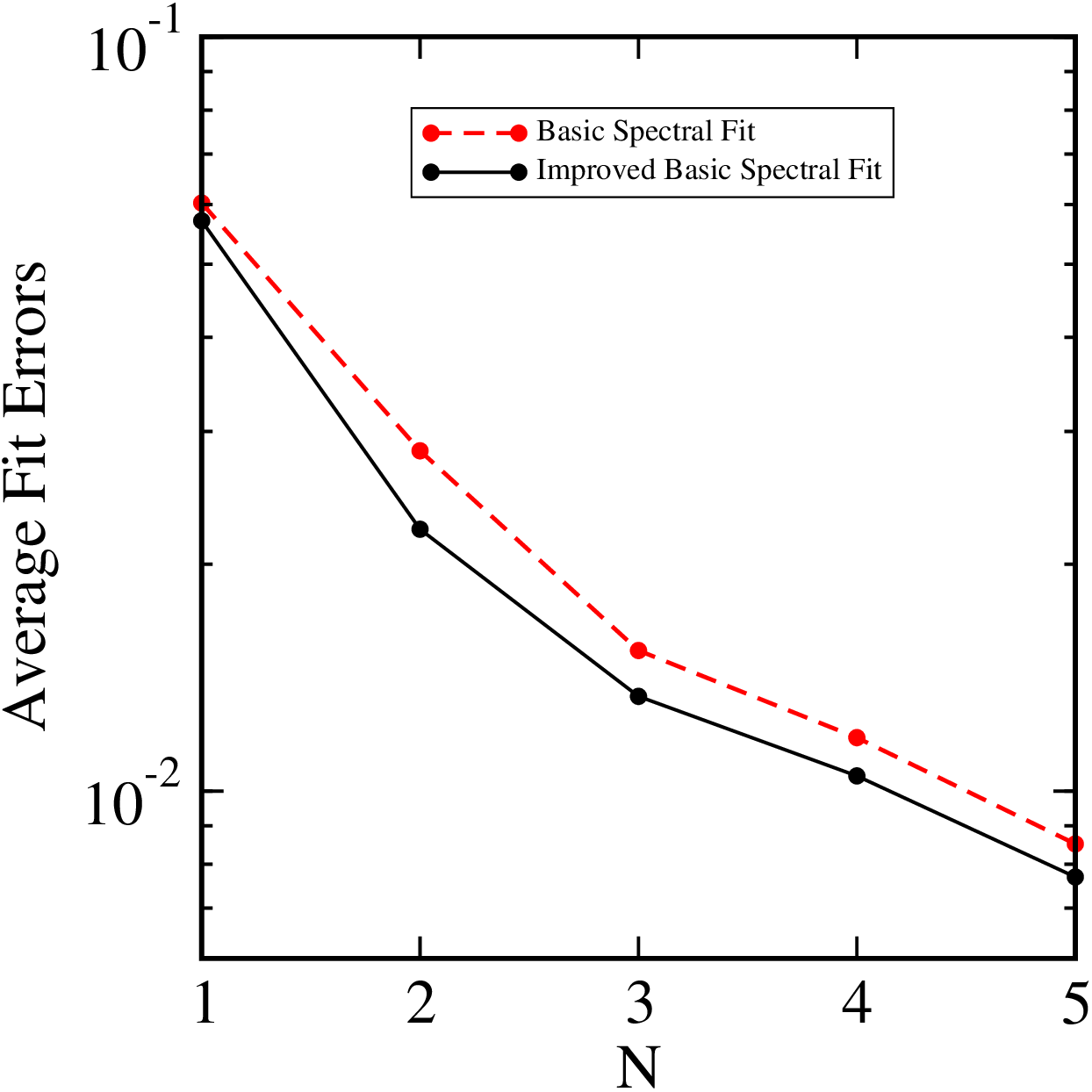}}
\caption{\label{f:AveragePFitErrors} Average errors of the
  $n$-parameter pressure-based basic and improved basic spectral fits
  are illustrated for 27 realistic neutron-star equations of state.}
\end{figure}

The accuracy and efficiency of these improved basic spectral
representations have been tested by computing best-fit models of the
27 (causal) theoretical neutron-star equations of state described
in~\cite{Lindblom2018}.  Figure~\ref{f:AveragePFitErrors} illustrates
the averages of these fitting errors over the set of test equations of
state for the $n=1,...,5$ basic and improved basic spectral
representations.  These results show that the improved basic spectral
representations, in addition to removing the unphysical phase
transition discontinuities at the matching point, are also somewhat
more accurate on average than the basic spectral representations with
the same number of adjustable parameters.

\section{Causal Spectral Representations}
\label{e:CausalRepresentations}

Spectral representations of the equation of state have also been
developed that ensure causality~\cite{Lindblom2018}.  These causal
representations are faithful in the sense that every causal equation
of state can be represented in this way, and every choice of spectral
coefficients in these representations produces a causal equation of
state.  These causal representations are based on spectral expansions
of the sound-speed function $\Upsilon(p)$, defined by
\begin{equation}
  \Upsilon(p)=\frac{c^2-v^2(p)}{v^2(p)},
  \label{e:UpsilonDev}
\end{equation}
where $v(p)$ is the sound speed of the material,
\begin{equation}
  v^2(p)=\left(\frac{d\epsilon(p)}{dp}\right)^{-1},
  \label{e:soundspeedDef}
\end{equation}
and $c$ is the speed of light.  The sound speed must be less than
the speed of light, $v(p)\leq c$, for any material that satisfies
causality, so $\Upsilon(p)\geq 0$ for these materials.

The equation of state, $\epsilon=\epsilon(p)$, determines
$\Upsilon(p)$ via Eqs.~(\ref{e:UpsilonDev}) and
(\ref{e:soundspeedDef}). Conversely, any $\Upsilon(p)$ on the domain
$p\geq p_0$ determines the equation of state, $\epsilon(p)$, on that
domain:
\begin{equation}
  \epsilon(p)=\epsilon(p_0)+\frac{p-p_0}{c^2}
  +\frac{1}{c^2}\int_{p_0}^p \Upsilon(p') dp',
\end{equation}
see~\cite{Lindblom2018}.  Therefore any $\Upsilon(p)\geq 0$ determines
a causal equation of state.
 
The causal spectral representations of the equation of state
introduced in~\cite{Lindblom2018} were based on the following spectral
expansions of $\Upsilon(p)$,
\begin{equation}
  \Upsilon(p) = \exp\left[\sum_{k} \lambda_k \Phi_k(p)\right],
\end{equation}
where $\Phi_k(p)$ are fixed spectral basis functions and $\lambda_k$
are spectral coefficients.  These spectral expansions determine an
$\Upsilon(p)$ that satisfies $\Upsilon(p)\geq 0$, and therefore a
causal equation of state.  If the $\Phi_k(p)$ are a complete set of
basis functions on a particular domain, then any $\Upsilon(p)$ can be
represented in this way on that domain.

These causal representations were shown to be efficient and accurate
ways to represent realistic neutron-star equations of state
in~\cite{Lindblom2018}.  In that study a collection of 27 different
causal nuclear-theory based neutron-star equations of state were used
to test the accuracy of these representations.  Best-fit spectral
expansions were prepared by adjusting the lowest-order spectral
coefficients, $\{\lambda_0,\lambda_1,...,\lambda_{n-1}\}$, to minimize
the differences between the exact and the spectral representation of
each equation of state.  These spectral expansions were quite
accurate, with average errors of $2-3\%$ for $n=2$ representations and
progressively smaller errors as the number of basis functions $n$ is
increased.  The spectral representations had systematically smaller
average errors than other popular piecewise-analytical representations
with the same number of adjustable parameters.  Therefore the
spectral-based representations were found to be both accurate and
efficient.

The spectral basis functions used in~\cite{Lindblom2018} for the
causal representations were the same as those used for the basic
representations in Eq.~(\ref{e:p-basis-def}).  It follows that the
causal spectral representations have $\Upsilon(p_0)=\exp(\lambda_0)$
at the point $(p_0,\epsilon_0)$ where the high-density spectral
representation matches onto the low-density equation of
state. Therefore the sound speed at this point in the causal spectral
representations is determined entirely by $\lambda_0$:
\begin{equation}
  v^2(p_0)=\frac{c^2}{1+\exp(\lambda_0)}.
  \label{e:upper-match_v}
\end{equation}
The sound speed at this matching point can also be determined from the
low-density equation of state in the usual way:
\begin{equation}
  v^2_0 = \left(\left.\frac{d\epsilon(p)}{dp}\right|_{p\uparrow
    p_0}\right)^{-1}.
  \label{e:lower-match_v}
\end{equation}
It follows that the sound speed (and hence the derivative of the
equation of state) will be continuous at this matching point if and
only if $v(p_0)=v_0$.  This agreement is possible if and only
if the spectral parameter $\lambda_0$ has the value,
\begin{equation}
  \lambda_0=\log\left[\frac{c^2-v^2_0}{v^2_0}\right].
  \label{e:lambda0}
\end{equation}

The $n$-parameter spectral representations described
in~\cite{Lindblom2018} use the $n$ lowest-order spectral parameters,
$\{\lambda_0,\lambda_1, ...,\lambda_{n-1}\}$ to determine the equation
of state.  The best possible match between the exact nuclear-theory
equations of state, and their spectral representations is achieved by
adjusting the values of these $n$ spectral parameters to minimize the
differences.  Similar to the basic spectral representations discussed
in Sec.~\ref{e:BasicRepresentations}, these optimal parameter choices
typically produce $\lambda_0$ that do not satisfy
Eq.~(\ref{e:lambda0}) exactly.  The graphs of the resulting best-fit
causal spectral representations in the neighborhood of the matching
point are similar to those shown in Fig.~\ref{f:spectral_apr1}, so
those graphs are not reproduced here. All the graphs show
$n$-dependent discontinuities at the matching point with the
low-density equation of state.  There is, however, one qualitative
difference between the causal graphs and those for the basic spectral
representations: In the causal spectral representations the
discontinuity in the slopes at the matching point are significantly
larger for the $n=1$ representations than that shown in
Fig.~\ref{f:spectral_apr1}.

\begin{figure}[b]
\centerline{\includegraphics[width=3in]{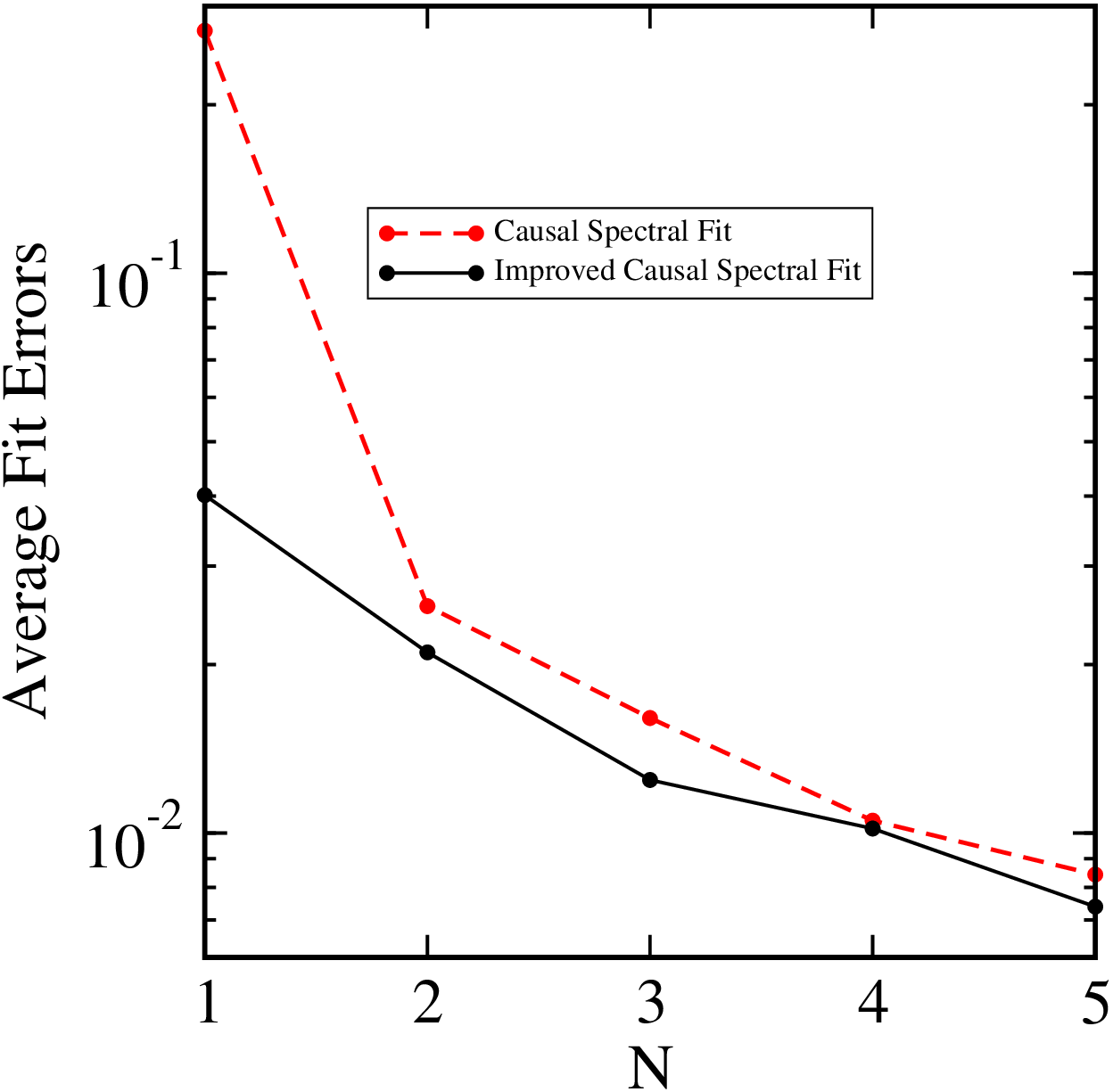}}
\caption{\label{f:AveragePCausalFitErrors} Figure illustrates the
  average errors of the $n$-parameter pressure-based causal and the
  improved causal spectral fits to 27 realistic neutron-star equations
  of state.}
\end{figure}
The causal spectral representations can easily be improved, however,
by setting $\lambda_0$ using Eq.~(\ref{e:lambda0}) to ensure
continuity of the sound speed at the matching point, and then
adjusting the $n$ parameters $\{\lambda_1,\lambda_2,...,\lambda_n\}$
instead of $\{\lambda_0,\lambda_1,...\lambda_{n-1}\}$.  The accuracy
and efficiency of these improved representations have been tested by
comparing the best-fit models for the 27 (causal) model neutron-star
equations of state described in~\cite{Lindblom2018}.
Figure~\ref{f:AveragePCausalFitErrors} illustrates the averages of
these fitting errors over the set of test equations of state for the
$n=1,...,5$ causal and improved causal representations.  These results
show that the improved causal spectral representations, in addition to
removing the unphysical phase transition discontinuities at the
matching point, are also more accurate on average than the causal
spectral representations with the same number of adjustable
parameters.  This improvement in accuracy is most pronounced for
representations with small numbers of basis functions.  The currently
available data from neutron-star observations is not accurate enough
or plentiful enough to allow very precise measurements of the equation
of state.  So having very accurate low-order representations should be
particularly useful at this time.

\section{Enthalpy Based Representations}
\label{s:EnthalpyRepresentations}

For some purposes it is more convenient to express the equation of
state, $\epsilon=\epsilon(p)$, in terms of the enthalpy of the
material, $h(p)=\int^p_0 dp'/\left[\epsilon(p')+p'\right]$, see
e.g.~\cite{Lindblom1992}.  The enthalpy version of the equation of
state is generally expressed as the pair of functions
$\epsilon=\epsilon(h)$ and $p=p(h)$.  Basic spectral representations
of these enthalpy-based equations of state have been given
in~\cite{Lindblom2010} and causal representations
in~\cite{Lindblom2018}.  Like their pressure-based counterparts, the
basic enthalpy-based representations are based on spectral expansions
of the adiabatic index $\Gamma(h)$, which then determines the equation
of state by quadratures (see~\cite{Lindblom2010}):
\begin{eqnarray}
p(h)&=&p_0 \exp\left[\int_{h_0}^h \frac{e^{h'}dh'}{\tilde \mu(h')}
\right],\label{e:PressueH}\\
\epsilon(h)&=& p(h)  \frac{e^h -\tilde \mu(h)}{\tilde \mu(h)},
\label{e:EnthalpyH}
\end{eqnarray}
where $\tilde \mu(h)$ is defined as,
\begin{eqnarray}
\tilde \mu(h) = \frac{p_0\, e^{h_0}}{\epsilon_0 + p_0} 
+ \int_{h_0}^h \frac{\Gamma(h')-1}{\Gamma(h')} e^{h'}dh'.
\label{e:TildeMuDef}
\end{eqnarray}
The constants $p_0$ and $\epsilon_0$ are defined by $p_0=p(h_0)$ and
$\epsilon_0=\epsilon(h_0)$ respectively.

Similarly the enthalpy-based causal representations are based on
spectral expansions of the sound-speed function $\Upsilon(h)$, which
also determines the equation of state by quadratures
(see~\cite{Lindblom2018}):
\begin{eqnarray}
p(h)&=&p_0+ \left(\epsilon_0\,c^2+p_0\right)   \int_{h_0}^h\hat\mu(h')\,dh',
\label{e:PressueH}\\
\epsilon(h)& =& -p(h)\,c^{-2}+\left(\epsilon_0+p_0\,c^{-2}\right)\hat\mu(h),
\label{e:EnergyH}
\end{eqnarray}
where $\hat\mu(h)$ is defined as, 
\begin{eqnarray}
\hat\mu(h) &=& \exp\left\{ \int_{h_0}^h \left[2+\Upsilon(h')\right]dh'\right\}.
\label{e:MuHDef}
\end{eqnarray}

As in the pressure-based cases discussed in
Secs.~\ref{e:BasicRepresentations} and \ref{e:CausalRepresentations},
the continuity of the derivative of the equation of state at the
matching point where $\epsilon_0=\epsilon(h_0)$ and $p_0=p(h_o)$ is
determined by the continuity of $\Gamma(h_0)$ or $\Upsilon(h_0)$ at
this point.  The spectral basis functions $\Phi_k(h)$ used in the
accuracy and efficiency studies in~\cite{Lindblom2010} and
\cite{Lindblom2018} are given by
\begin{equation}
  \Phi_k(h)=\left[\log\left(\frac{h}{h_0}\right)\right]^k.
\end{equation}
This basis has the property that $\Phi_0(h_0)=1$ and $\Phi_k(h_0)=0$
for $k>0$. Therefore the expressions for $\Gamma(h_0)=\exp(\gamma_0)$
and $\Upsilon(h_0)=\exp(\lambda_0)$ at the matching point are
identical to those for the pressure-based representations.
Consequently the conditions needed to ensure the enthalpy-based
spectral equations of state match the low-density equation of state
with continuous derivatives are identical to those for the
pressure-based representations: Eqs.~(\ref{e:gamm0def}) and
(\ref{e:lambda0}) respectively.

\begin{figure}[tb]
\centerline{\includegraphics[width=3in]{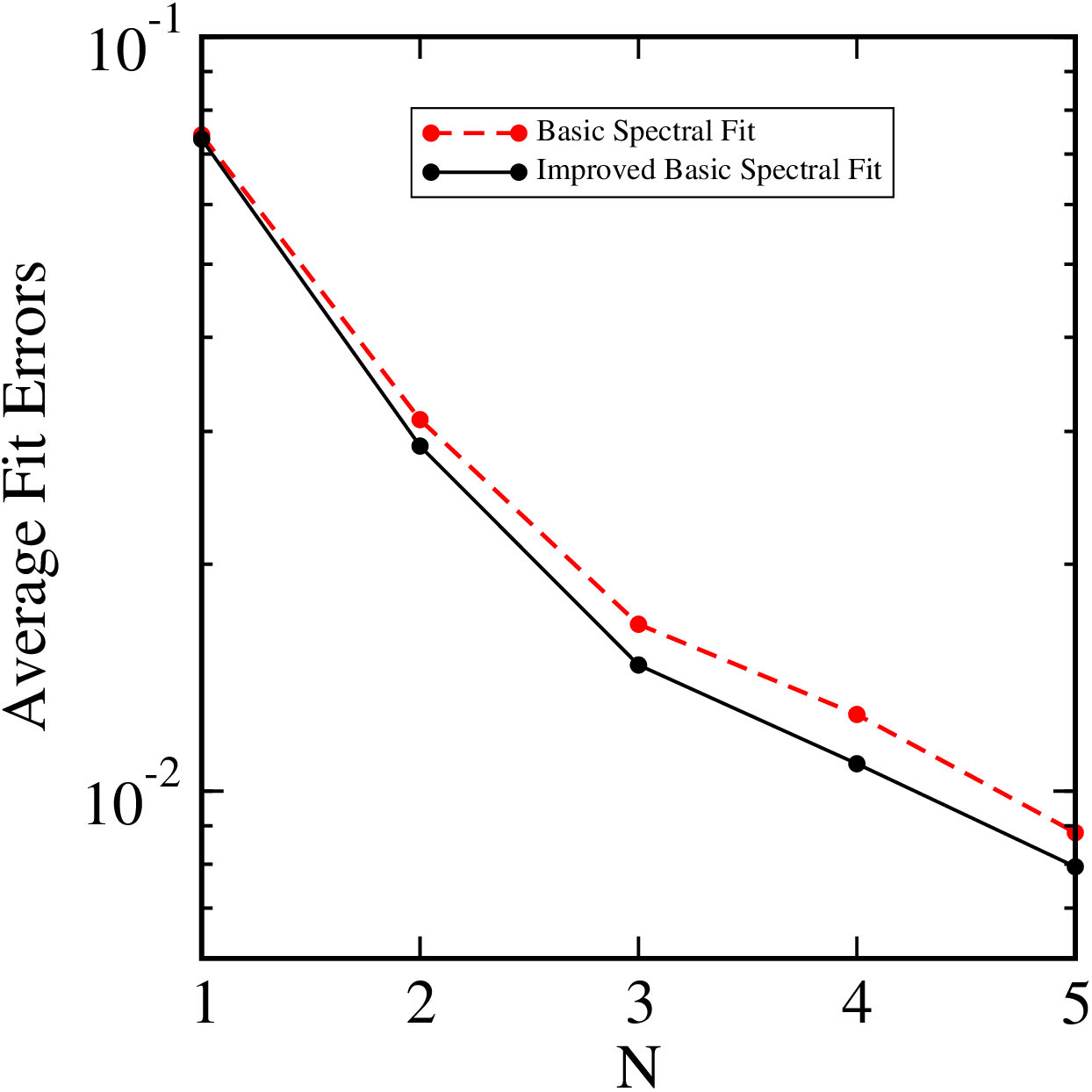}}
\caption{\label{f:AverageHFitErrors} Average errors of the
  $n$-parameter enthalpy-based basic and improved basic spectral fits
  are illustrated for 27 realistic neutron-star equations of state.}
\end{figure}
\begin{figure}[b]
\centerline{\includegraphics[width=3in]{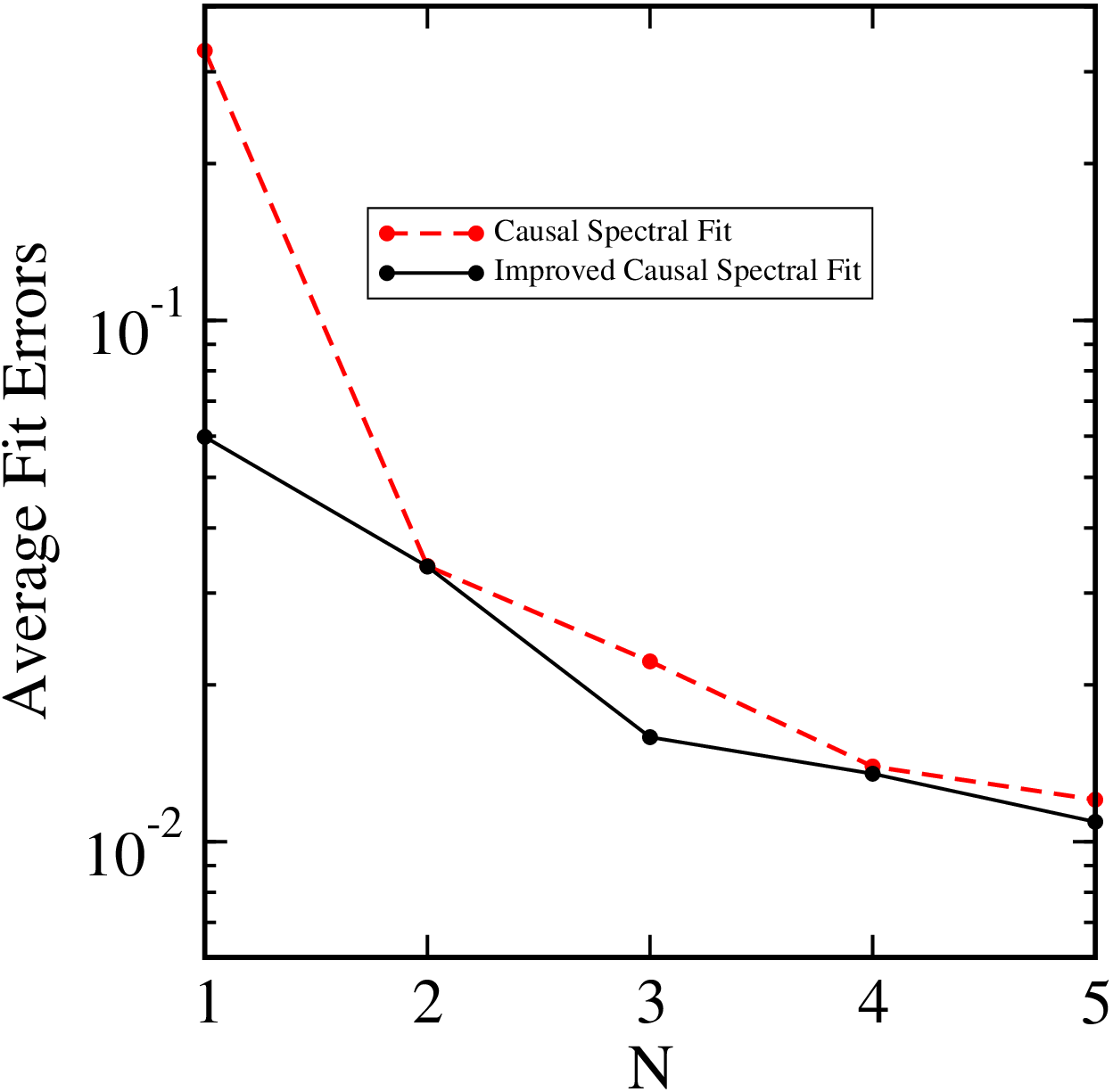}}
\caption{\label{f:AverageHCausalFitErrors} Average errors of the
  $n$-parameter enthalpy-based causal and improved causal spectral
  fits are illustrated for 27 realistic neutron-star equations of
  state.}
\end{figure}

The enthalpy-based spectral expansions can therefore be improved using
the same method used for the pressure-based representations in
Secs.~\ref{e:BasicRepresentations} and \ref{e:CausalRepresentations}.
Instead of adjusting the parameters
$\{\gamma_0,\gamma_1,...,\gamma_{n-1}\}$ or
$\{\lambda_0,\lambda_1,...,\lambda_{n-1}\}$ to obtain the optimal
$n$-parameter representations, the lowest-order parameters, $\gamma_0$
or $\lambda_0$, are fixed using Eqs.~(\ref{e:gamm0def}) or
(\ref{e:lambda0}), while the $n$ parameters
$\{\gamma_1,\gamma_2,...,\gamma_{n}\}$ or
$\{\lambda_1,\lambda_2,...,\lambda_{n}\}$ are adjusted for the best
fits.  The average errors in the best fits to 27 realistic
neutron-star equations of state are illustrated in
Fig.~\ref{f:AverageHFitErrors} for the basic and the improved basic
enthalpy-based representations, and in
Fig.~\ref{f:AverageHCausalFitErrors} for the causal and the improved
causal representations.  As in the pressure-based representations, the
accuracies of the improved enthalpy-based representations for a given
value of $n$ are systematically better than the original
representations.
%


\acknowledgments This research was supported in part by NSF grant
2012857 to the University of California at San Diego.

\vfill\eject
\bibstyle{prd} 
\bibliography{../References/References}
\end{document}